\documentclass[12pt]{iopart}

\usepackage{graphicx}
\usepackage{verbatim}
\usepackage{bm}
\usepackage{lineno}

\newcommand{\beq}{\begin{equation}}
\newcommand{\eeq}{\end{equation}}
\newcommand{\M}{\mathcal{M}}

\begin{document}

\title[Interlayer Hebbian Plasticity Induces First-Order Transition in Multiplex Networks]{Interlayer Hebbian Plasticity Induces First-Order Transition in Multiplex Networks}

\author{Ajay Deep Kachhvah$^1$, Xiangfeng Dai$^{2,3}$, Stefano Boccaletti$^{3,4,5}$ and Sarika Jalan$^{*1,6}$}
\address{$^1$ Complex Systems Lab, Indian Institute of Technology Indore - Simrol, Indore - 453552, India}
\address{$^2$ School of Mechanical Engineering and Center for OPTical IMagery Analysis and Learning (OPTIMAL), Northwestern Polytechnical University, Xi'an, Shaanxi, 710072, China}
\address{$^3$ Unmanned Systems Research Institute, Northwestern Polytechnical University, Xi'an, Shaanxi, 710072, China}

\address{$^4$ CNR - Institute of Complex Systems, Via Madonna del Piano 10, I-50019 Sesto Fiorentino, Italy}
\address{$^5$ Moscow Institute of Physics and Technology (National Research University), 9 Institutskiy per., Dolgoprudny, Moscow Region, 141701, Russian Federation}

\address{$^6$ Center for Theoretical Physics of Complex Systems, Institute for Basic Science (IBS), Daejeon 34126, Korea}

\ead{*Corresponding Author:sarika@iiti.ac.in}
\vspace{10pt}

\begin{abstract}
Adaptation plays a pivotal role in the evolution of natural and artificial complex systems, and in the determination of  their functionality. Here, we investigate the impact of adaptive inter-layer processes on intra-layer synchronization in multiplex networks. The considered adaptation mechanism is governed by a Hebbian learning rule, i.e., the link weight between a pair of interconnected nodes is enhanced if the two nodes are in phase. Such adaptive coupling induces an irreversible first-order transition route to synchronization accompanied with a hysteresis. We provide rigorous analytic predictions of the critical coupling strengths for the onset of synchronization and de-synchronization, and verify all our theoretical predictions by means of extensive numerical simulations.
\end{abstract}

%
\vspace{2pc}
%
%
%
%
\section{Introduction}
The study of synchronization, a collective motion of initially un-identical  units interacting on network structures has provided a deeper understanding on the underlying processes (and the nature of transition) taking place on a wide range of physical and biological systems~\cite{Pikovsky2003}. Second-order-like transitions in systems of phase oscillators are frequent in nature, whereas abrupt, discontinuous, first-order-like transitions are not so common. However, a plethora of recent studies has established that first-order-like transitions can be achieved in networked oscillators under certain circumstances inducing frustration mechanisms preventing synchronization between connected oscillators, and thus blocking the formation of giant clusters during the transition. Such a transition (called explosive synchronization) usually involves an abrupt formation of the giant synchronized cluster, and then its irreversible abrupt de-synchronization, yielding a hysteresis loop. The hypersensitivity or abruptness arisen due to a minuscule change in the interaction strength makes this process unmanageable and calamitous in many circumstances. Examples are, for instance, blackouts in the power grid~\cite{Buldyrev2010}, breakdown of the internet~\cite{Huberman1997}, episodes of Fibromyalgia chronic pain~\cite{Lee2018} and epileptic seizures~\cite{Adhikari2013} in the human brain. The explosive synchronization (ES) transition is shown to emerge in networked oscillators with microscopic correlation between their frequency and network's structural property such as degree~\cite{Gardenes2011} and coupling strength~\cite{Zhang2013}, or by inclusion of inertia~\cite{Tanaka1997, Olmi2014, Gupta2014} and noise~\cite{Gupta2014, Bonilla1992}, or by the presence of a fraction of adaptively coupled oscillators~\cite{Zhang2015, Danziger2016, Dai2020, Khanra2018, Ling2020, Frolov2020}, mean-field diffusion~\cite{Verma2017}, traffic processes~\cite{Chen2020}, the presence of nearest-neighbor competitive interaction or symmetry-breaking interaction~\cite{Shamik2020} and anti-Hebbian adaptation of link weight~\cite{AvalosG2018}.

Despite being a very useful framework to investigate phenomena such as synchronization and percolation, isolated networks are often unable to precisely represent the behavior of complex systems involving different types of interactions among the same set of interacting elements. Multilayer or multiplex networks are the right candidates to map such systems, as they comprise different types of connections or processes among a common set of nodes, interconnected through different layers~\cite{Boccaletti2014,Wang2020,Rybalova2020,Majhi2019,Sawicki2018}. The advent of multiplex network has made it possible to investigate the impact of one type of process (layer) on other interdependent processes (other layers) such as synchronization, percolation, epidemic spreading, etc. In the same breath, a few techniques have been employed inducing ES in one or all dynamical layers. For instance, ES is common in adaptively coupled interdependent layers~\cite{Zhang2015}, in random walker dynamics~\cite{Nicosia2017}, in presence of a delay~\cite{Ajaydeep2019} or interlayer adaptation through order parameter of the layers ~\cite{Kumar2020}.

The mechanism of adaptation plays a crucial role in the development and function of many natural and artificial systems. In neuroscience, most of the experimental findings suggest that synaptic adaptation between neurons is the basis of learning and long-term memory~\cite{Shimizu2000,Abbott2000}. First proposed by Hebb~\cite{Hebb1949} and later supported by experimental evidences~\cite{Abbott2000,Markram1997,Zhang1998}, such adaptation consists in the fact that the synaptic coupling between two neurons is strengthened or weakened if the two neurons are simultaneously firing or not depending on the pinpoint relative timing of presynaptic and postsynaptic spikes. If the relative spike timing is coded in terms of the phases of the oscillators, a neural network then can be delineated as a network of phase oscillators. Thus, the adaptive evolution of the neural network occurs through the alterations of synaptic connections between neurons. The adaptation of connection-weights following such Hebbian learning mechanism leads to the occurrence of cluster states~\cite{Niyogi2009,Aoki2009,Berner2020} or meso-scale structures~\cite{Gutierrez2011,Pitsik2018} underlying synchronization in complex networks.  

In this paper, we investigate a multiplex framework inspired by such Hebbian adaptive learning rule. In our network, the weights of the links between interconnected layers are adaptive and governed by the instantaneous phase-difference of the interconnected nodes. Strikingly, the Hebbian learning rule divides the population of interlayer links into weights and anti-weights for low intra-layer interaction strength. The existence of the inhibitory interlayer weights drives the multiplexed layers adopt the first-order transition route to synchronization accompanied with a hysteresis. The Hebbian learning mechanism also provides a great amount of control over the abruptness and the width of associated hysteresis by means of learning parameters. It is further revealed that the critical coupling strength for the onset of synchronization does show explicit dependence on the learning parameter while that for the onset of desynchronization remains independent of it. The numerical assessments of both the critical coupling strength have shown a good match with their respective analytical predictions. The proposed recipe based on Hebbian adaptation is capable of triggering  first-order transitions in all homogeneous networks.

\section{\label{sec:methods}Model}
Let us start with investigating how the Hebb's neural learning mechanism governing the inter-layer  weight (strength) affects the phase transition in the multiplexed layers. In order to achieve this, we consider a multiplex network composed of two layers of the same size $N$. The dynamics of nodes in each layer is governed by the Kuramoto model~\cite{Kuramoto1984}. The inter-layer link weight $D^{[ii]}$ between node $i$ in a layer and its counterpart in the other layer is adaptive in nature. Hence, the evolution of the phase oscillators is governed by
\begin{eqnarray}\label{eqn1}
   	\dot\theta^{[i]}_1 &= \omega^{[i]}_1+ \lambda_1 \sum_{j=1}^{N} A^{[ij]}_1 \sin(\theta_1^{[j]}-\theta_1^{[i]}) + D^{[ii]} \sin(\theta_2^{[i]}-\theta_1^{[i]}), \nonumber  \\
    \dot\theta^{[i]}_2 &= \omega^{[i]}_2+ \lambda_2 \sum_{j=1}^{N} A^{[ij]}_2 \sin(\theta_2^{[j]}-\theta_2^{[i]}) + D^{[ii]} \sin(\theta_1^{[i]}-\theta_2^{[i]}),
\end{eqnarray}
where $i=1,...,N$, subscripts $1$ and $2$ stand for the two distinct layers, and $\lambda$ represents intra-layer coupling strength, here $\lambda_1=\lambda_2=\lambda$. The instantaneous phase of the $i^{th}$ node is denoted by $\theta_l^{[i]}(t)$ and its natural frequency $\omega_l^{[i]}$ follows uniform or unimodal distribution $g(\omega_l)$. The intra-layer connectivity between the nodes following a network topology is encoded in the adjacency matrix $A_l$ such that $A_l^{[ij]}=1$ $(0)$ if $i^{th}$ and $j^{th}$ nodes are connected (disconnected). The dynamically adaptive weight $D^{[ii]}$ of an inter-layer link between each pair of interconnected (mirror) nodes in the two layers is determined by the following Hebbian learning rule
\begin{eqnarray} \label{eqn2}
	\dot D^{[ii]}=\varepsilon[\alpha\cos(\theta^{[i]}_2-\theta^{[i]}_1)-D^{[ii]}],
\end{eqnarray}
\begin{figure}[t!]
	\begin{center}
	\includegraphics[height=7cm,width=10cm]{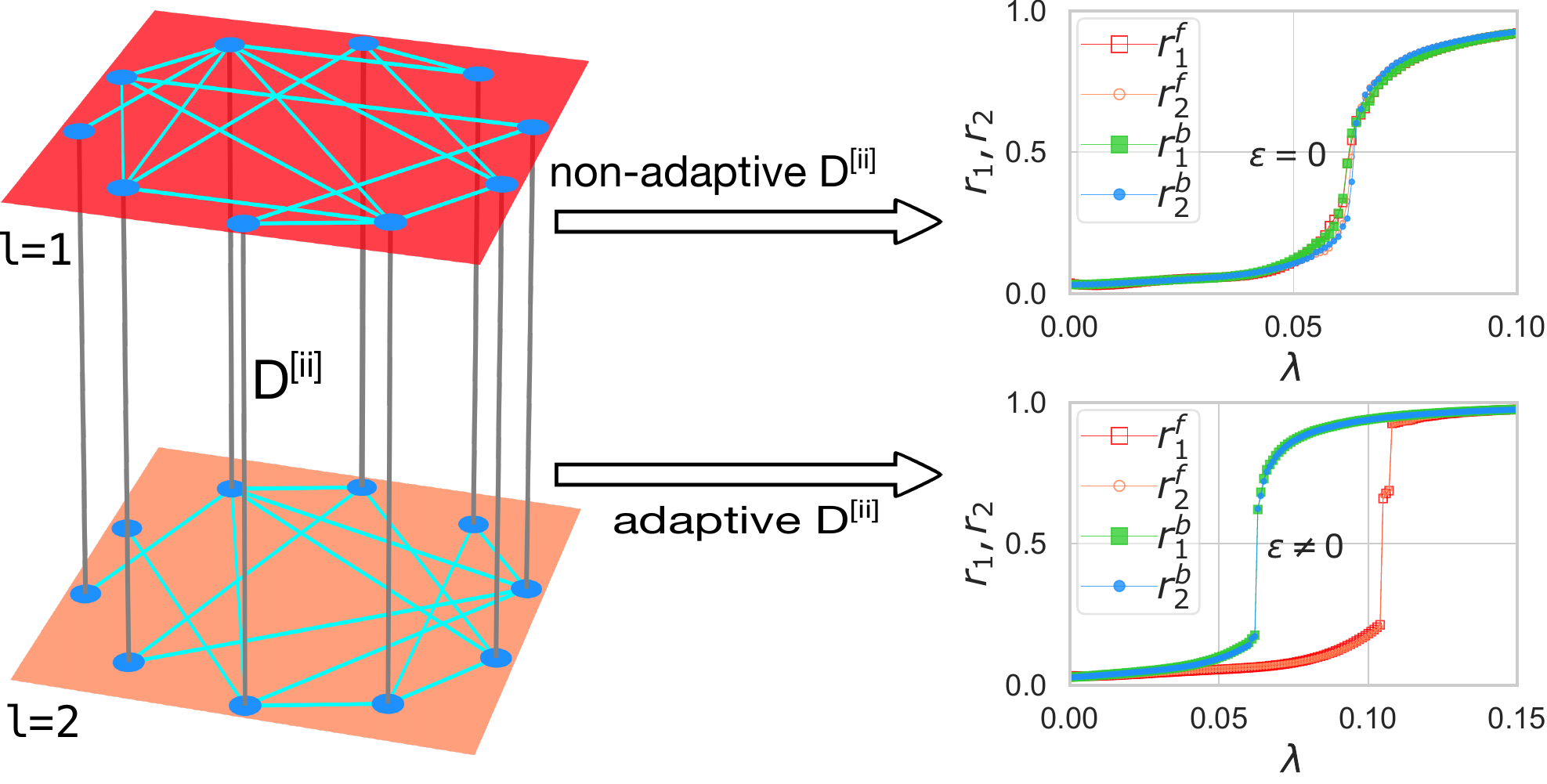}\\
	\vspace{-0.5cm}
	\caption{{\bf Schematic representation of the duplex network} ($l=1,2$) having non-adaptive or adaptive inter-layer coupling between interconnected nodes. The non-adaptive case (learning rate $\varepsilon=0$) leads to a continuous transition, whereas the Hebbian learning rule ($\varepsilon\neq0$) yields a discontinuous route to synchronization and then an irreversible route to desynchronization. $r^f$ and $r^b$ denote order parameter in the forward and backward continuation of coupling strength $\lambda$, respectively.}
	\end{center}
	\label{figure1}
\end{figure}
where $\alpha\in[0,1]$ is a factor which amplifies the amount of learning if the two nodes are synchronized and $\varepsilon\in[0,1]$ is the learning rate. The $D^{[ii]}$ in Eq.\ref{eqn2} prevents the inter-layer coupling weight from increasing or decreasing without bounds.
Hence, the supra-adjacency matrix of the multiplex network is adaptive, and includes un-weighted intra-layer links and adaptive weighted inter-layer links:
\begin{eqnarray}
\M\rm(t)=\left(
\begin{array}{cc}
A_1  & D\rm(t)I \\
D\rm(t)I & A_2 \\
\end{array}
\right),
\end{eqnarray}
where $I$ is the identity matrix.

To track the level of coherence in the system, we define the global order parameter $r_l$ for a layer $l$ in terms of the average phase $\psi_l$ as
\begin{equation}\label{eqn3}
    r_l(t)e^{\imath \psi_l} = \frac{\sum_{i=1}^{N}k_l^{[i]}e^{\imath\theta_l^{[i]}}}{\sum_{i=1}^{2N}k_l^{[i]}}.
\end{equation}
$r=1$ represents a completely synchronous state, while $r=0$ implies total incoherence. In a similar way, one can define in-phase (one-cluster) global order parameter for the entire multiplex network as
\begin{equation}\label{eqn4a}
    R_1(t)e^{\imath \Psi_l} = \frac{\sum_{i=1}^{2N}k_l^{[i]}e^{\imath\theta_l^{[i]}}}{\sum_{i=1}^{2N}k_l^{[i]}}.
\end{equation}
Furthermore, to capture the degree of anti-phase (two-cluster) global synchronization, a new global order parameter $R_2$ (dipole moment of the distribution of anti-phases) \cite{Niyogi2009, Karimian2019} is defined as follows
\begin{eqnarray}\label{eqn4b}
    {R_2} = {|R-R_1|},
\end{eqnarray}
where
\begin{eqnarray}
    R(t)e^{\imath \Phi_l} = \frac{\sum_{i=1}^{2N}k_l^{[i]}e^{2\imath\theta_l^{[i]}}}{\sum_{i=1}^{2N}k_l^{[i]}}.\nonumber
\end{eqnarray}
Here, $R$ measures both the one-cluster and the two-cluster synchronization. Hence in order to determine the degree of two-cluster synchronization, the term $R^{'}$ is adjusted in Eq.\ref{eqn4b} for one-cluster synchronization.
\begin{figure*}[t!]
	\begin{center}
	\includegraphics[height=5cm,width=8cm]{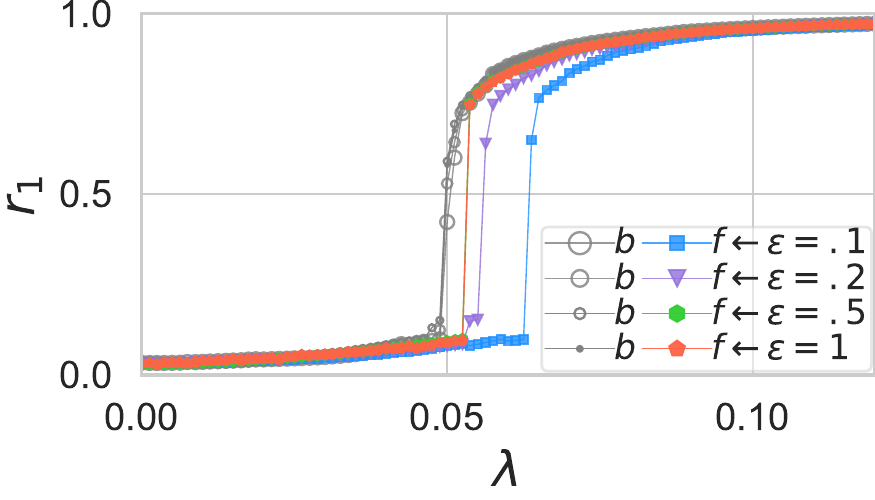}\\
	\vspace{-0.5cm}
	\caption{{\bf Adaptive multiplexing leads to ES.} Order parameter for forward and backward transitions ($r_l^f$ and $r_l^b$) for either multiplexed ER layer (since the two layers synchronize simultaneously) corresponding to $\alpha=0.5$ and different values of $\varepsilon$. The network parameters are $N=1,000$ and $\langle k\rangle=10$ and $\gamma=0.5$.}
	\label{figure2}
	\end{center}
\end{figure*}

\section{Results}
We consider a multiplex network made of two Erd\"{o}s-R\'{e}nyi (ER) random networks~\cite{Erdos1960} having average intra-layer connectivity $\langle k_1\rangle=\langle k_2\rangle=10$. The nodes in both layers are assigned initial phases and natural frequencies drawn from a uniform random distribution such that $\theta^{[i]}\in[0,2\pi)$ and $\omega^{[i]}\in[-\gamma,\gamma)$ where $\gamma=0.5$ unless otherwise stated, respectively. The initial $N$ values of $D^{[ii]}(t)$ are selected as $D^{[ii]}(t{=}0)=d^{[i]}/N$, where $d^{[i]}$ is the number of inter-layer links a node in one layer can have with the nodes in other layer. Here we adopt the simplest form of a multilayer network, i.e. $d^{[i]}=1,\forall i$.

First, we investigate the impact of the learning rate $\varepsilon$ on intra-layer synchronization in the multiplexed layers.
Fig.\ref{figure2} illustrates the behavior of the order parameter (for both forward and backward transitions) for the two layers as a function of coupling strength $\lambda$ for different values of learning rate $\varepsilon$ sustained with a fixed choice of learning factor $\alpha$. It is found that the two layers undergo a first-order transition (ES) for different values of $\varepsilon$. It is also unveiled that the backward critical coupling strength $\lambda_c^b$ is independent of the rate $\varepsilon$. However, the forward critical coupling strength $\lambda_c^f$ at first decreases with the increase in $\varepsilon$, then no further change is observed for higher values ($0.5-1$) of $\alpha$. Hence slower learning rates $\varepsilon$ yield slightly wider hystereses.

\begin{figure}[t!]
	\begin{center}
	\includegraphics[height=7cm,width=14cm]{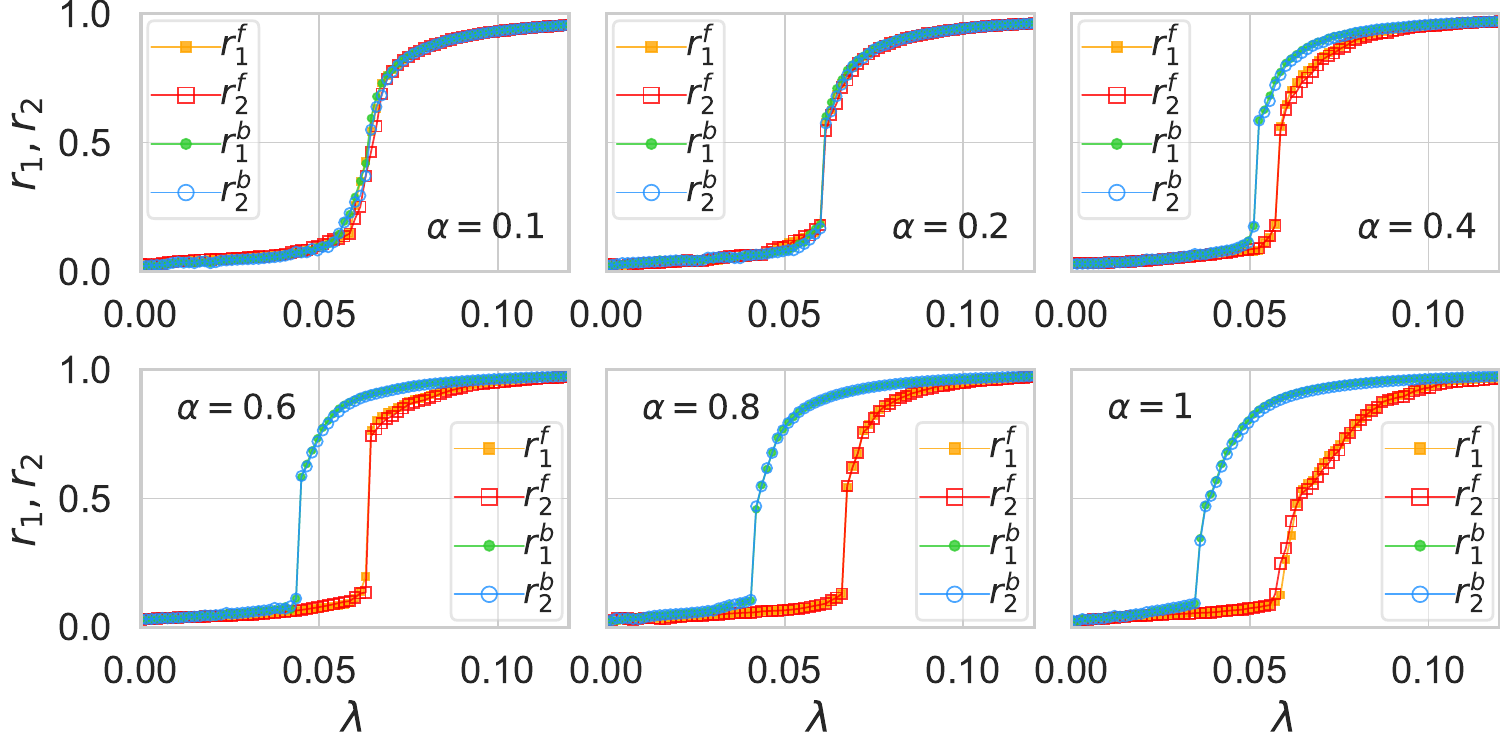}\\
	\vspace{-0.5cm}
	\caption{{\bf Impact of $\alpha$:} Order parameters $r_l^f$ (forward) and $r_l^b$ (backward) for the two multiplexed ER layers. $\varepsilon=0.1$ and different values of $\alpha$. Network parameters are the same as in the caption of Fig.\ref{figure2}.}
	\label{figure3}
	\end{center}
\end{figure}
Next, the impact of the amplification factor $\alpha$ with a fixed choice of $\varepsilon$ is reported in Fig.\ref{figure3}. The two layers follow a first-order transition with significantly different hysteresis width for different values of $\alpha$. At very low $\alpha$, a second-order transition is observed. With the increase in $\alpha$, the forward critical $\lambda_c^f$ significantly increases while the backward critical coupling $\lambda_c^b$ remains independent on $\alpha$ as well. Thus, each increase in $\alpha$ leads to a broader hysteresis width.

\begin{figure}[t]
	\begin{center}
	\includegraphics[height=12cm,width=15cm]{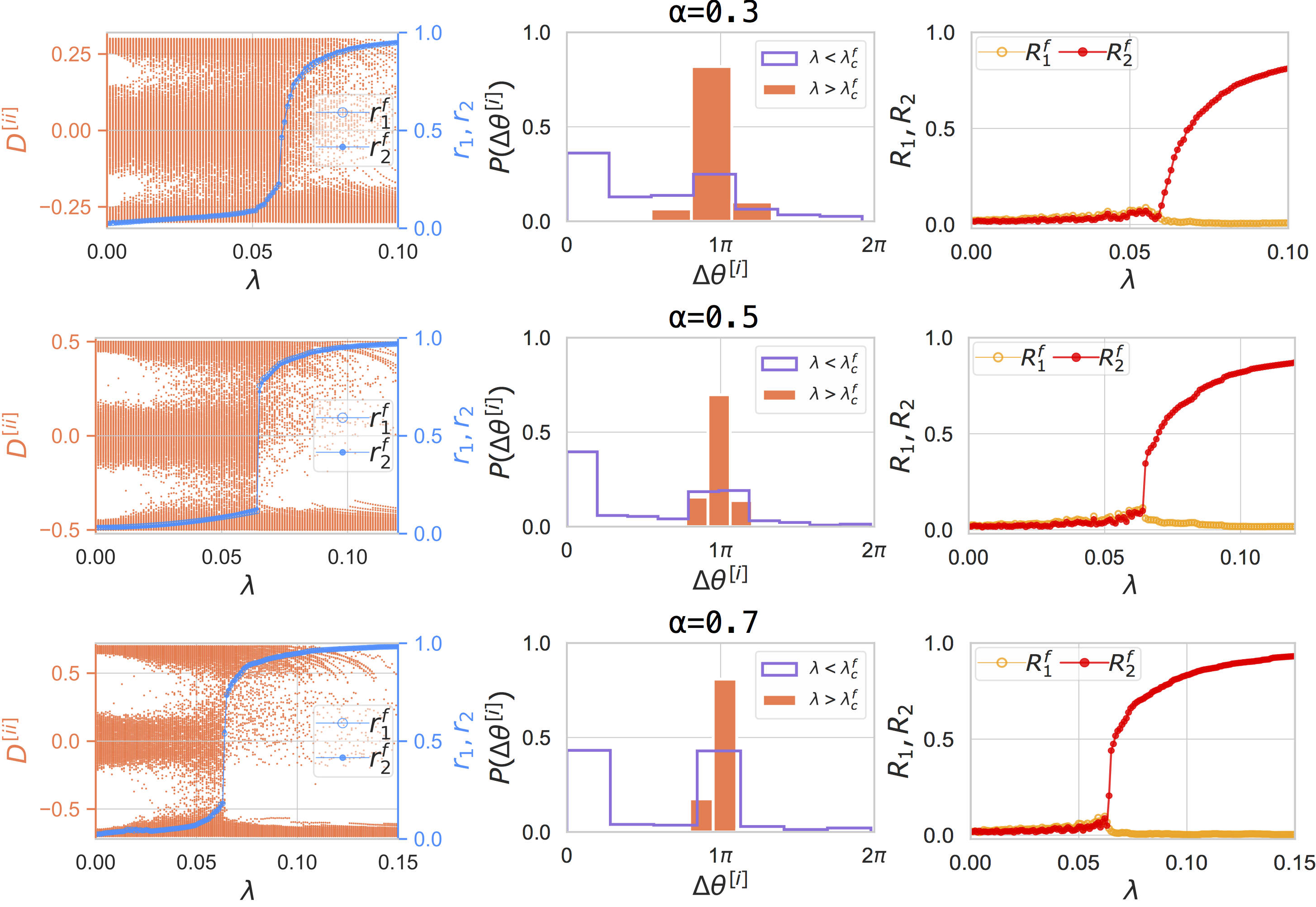}\\
	\vspace{-0.5cm}
	\caption{{\bf Mechanism behind ES} $D^{[ii]}$ as a function of $\lambda$, the distribution $P(\Delta\theta^{[i]})$ of $\Delta\theta^{[i]}=|\theta_2^{[i]}-\theta_1^{[i]}|$ belonging to the incoherent and coherent state, and in-phase ($R_1$) and anti-phase ($R_2$) global order parameter as a function of $\lambda$ for the multiplex network made of two ER layers. Results are produced for $N{=}1,000$ nodes, $\varepsilon=0.1$ and $\gamma=0.5$ and different values of $\alpha$.}
	\label{figure4}
	\end{center}
\end{figure}

To gather information at a microscopic level on the origin of the first-order transition, we investigate the behavior of $D^{[ii]}$, the distribution $P(\Delta\theta^{[i]})$ of $\Delta\theta^{[i]}=|\theta_2^{[i]}-\theta_1^{[i]}|$, in phase ($R_1$) and anti-phase ($R_2$) global order parameter in Fig.\ref{figure4}.
The first column of Fig.\ref{figure4} reports the final state of $D^{[ii]}$ in the forward continuation of $\lambda$ for different values of $\alpha$ at a fixed rate $\varepsilon$. It is apparent that the stationary values of $D^{[ii]}=D^{[ii]}(t)$ are bounded in the interval $[-\alpha, \alpha]$. In the incoherent state ($\lambda<\lambda_c^{f,b}$), the inter-layer link population is mainly divided into two clusters of anti-weights, namely $-\alpha$ and $\alpha$. Besides, there exists a neutral cluster around $D^{[ii]}\rightarrow 0$ containing a small fraction of link population, which increases with the decrease in the value of $\alpha$. Interestingly, $D^{[ii]}$ and $-D^{[ii]}$ populations are almost equal in size. On the contrary, the inter-layer link population tends to converge to $-\alpha$ in the coherent state ($\lambda>\lambda_c^{f,b}$). The inter-layer link population having $D^{[ii]}\rightarrow-\alpha$ in the incoherent state originates a frustration at the respective interconnected end nodes. The triggered frustration at the end nodes of the inhibited (subjected to $-\alpha$) inter-layer links curbs the formation of the largest synchronous cluster in their respective layers until a threshold for $\lambda$ is reached. The larger the magnitude $-\alpha$, the stronger the triggered frustration. Since a large value of $-\alpha$ imparts a stronger inhibition during the forward continuation of $\lambda$, a stronger and stronger $\lambda$ is required for the abrupt formation of the largest synchronous cluster with each increase in $\alpha$. Thus, the onset of first-order transition is witnessed at a larger forward critical coupling $\lambda_c^f$ with each increase in $\alpha$.

In the second column of Fig.\ref{figure4}, we study the distribution $P(\Delta\theta^{[i]})$ of $\Delta\theta^{[i]}=|\theta_2^{[i]}-\theta_1^{[i]}|$, the difference between phases of the interconnected nodes for different values of $\alpha$ while keeping $\varepsilon$ fixed. It unveils that in the incoherent state ($\lambda<\lambda_c^f$ ) belonging to an intermediate or higher value of $\alpha$, two phase-clusters in each layer exist: one corresponding to $\theta_2^{[i]}=\theta_1^{[i]}$ and the other to  $|\theta_2^{[i]}-\theta_1^{[i]}|=\pi$. In addition, the neutral cluster population remains distributed between these two clusters, whose population increases with the decrease in $\alpha$. Hence, each multiplexed layer stays in a bi-clusters state in the incoherent state. Nevertheless, the $P(\theta^{[i]})$ in the coherent state ($\lambda>\lambda_c^f$) unveils a sharp single peaked (unimodal) distribution centered at $|\theta_2^{[i]}-\theta_1^{[i]}|=\pi$ for any value of $\alpha$. It implies that both layers are locked to two different phases at a mutual difference of $\pi$, i.e., the multiplex network comprises two anti-phase layers in the coherent state.

The third column of Fig.\ref{figure4} illustrates $R_1$ (Eq.\ref{eqn4a}) and $R_2$ (Eq.\ref{eqn4b}) for different values of $\alpha$. In the incoherent state one has that $R_1\rightarrow0$ and $R_2\rightarrow0$, as there exists two anti-phase clusters with $\theta^{[i]}=0$ and $\theta^{[i]}=\pi$ in each layer. In the coherent state, however, there exists a single cluster centered at $\theta^{[i]}=\pi$, hence still one ha $R_1\rightarrow0$, while $R_2$ displays an abrupt jump at the critical coupling strength. The height of the abrupt jump for $R_2$ increases as $\alpha$ increases.
	\begin{figure}[t!]
	\begin{center}
	\begin{tabular}{cc}
		\includegraphics[height=6cm,width=8cm]{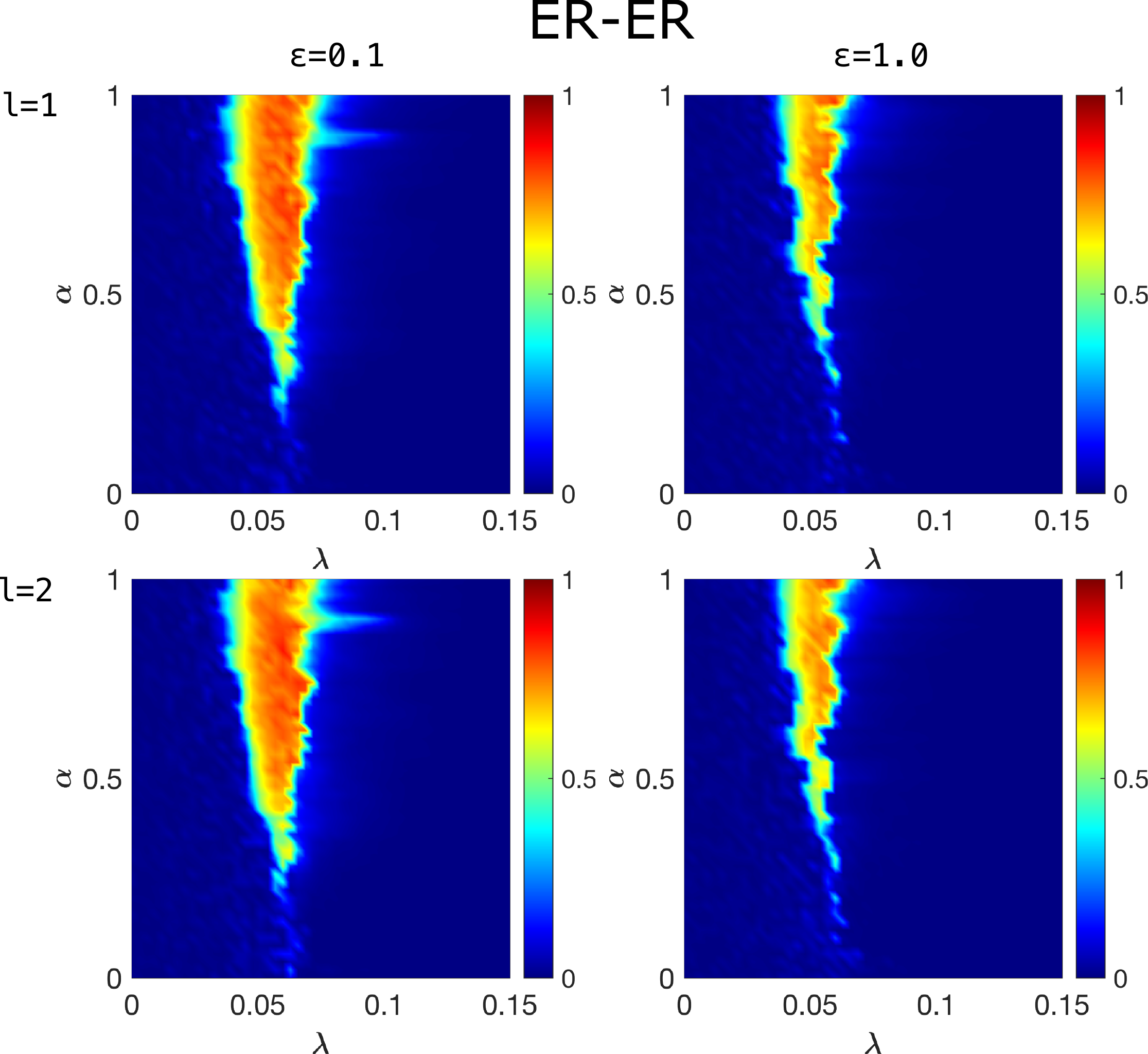}&
		\includegraphics[height=6cm,width=8cm]{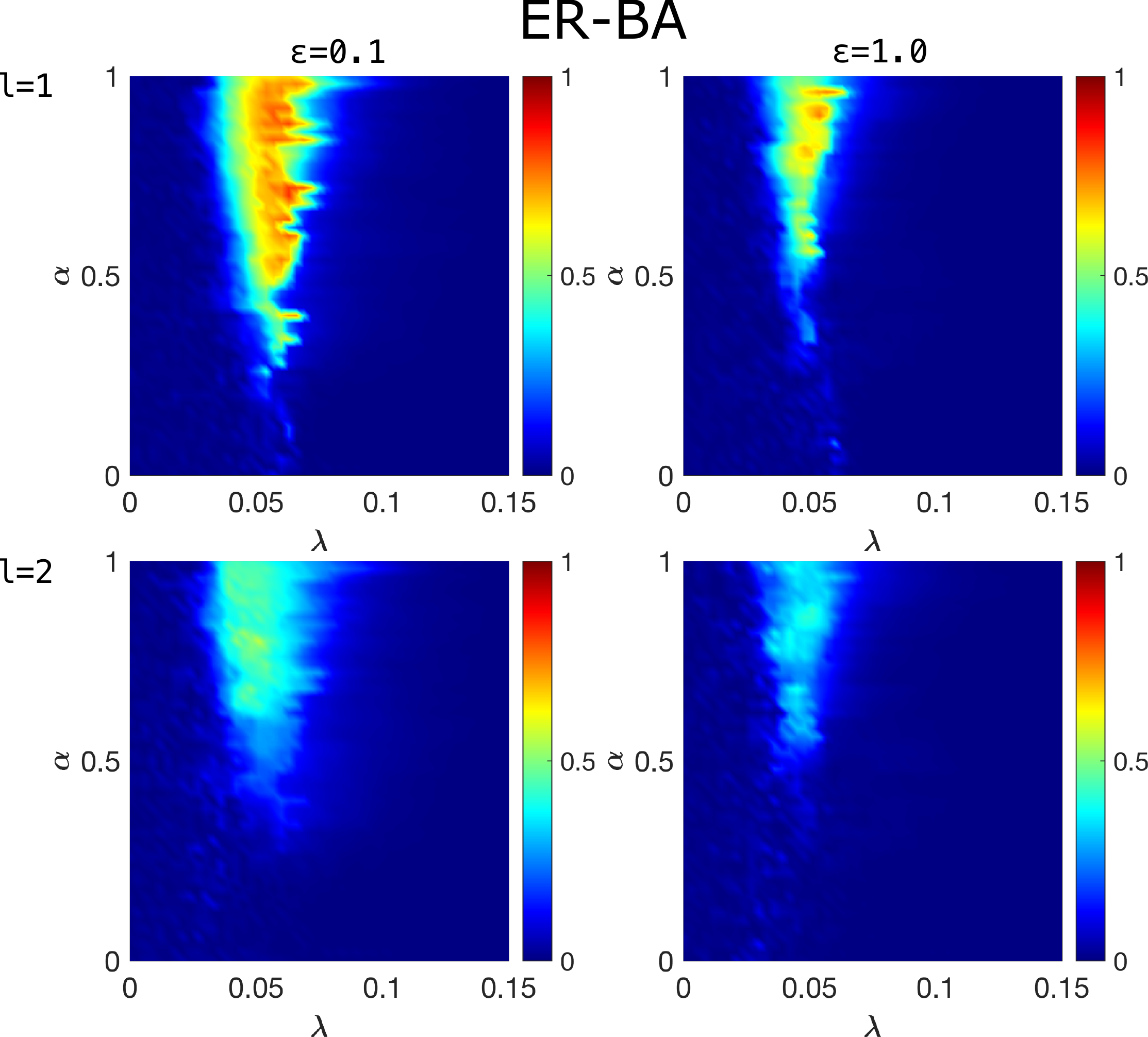}\\
	\end{tabular}{}
	\vspace{-0.5cm}
	\caption{Phase diagrams $\alpha-\lambda$ of hysteresis width $|r_l^f-r_l^b|$, where $l\in [1,2]$, corresponding to two different values of $\varepsilon$  ($N=1,000$, $\gamma=0.5$). Data refers to pairs of a variety of network topology, namely ER-ER and ER-BA. Note that layer $l=2$ denotes BA network for ER-BA configuration.}
	\label{figure5}
	\end{center}
\end{figure}

Now, since $\dot D^{[ii]}=0$ (from Eq.\ref{eqn2}) in the stationary state, one obtains (for $\varepsilon\ne0$)
\begin{eqnarray}\label{eqn4}
	D^{[ii]}=\alpha\cos[\theta^{[i]}_2-\theta^{[i]}_1]=\alpha\cos(\Delta\theta^{[i]}).
\end{eqnarray}
Further, it is revealed from Fig.~\ref{figure4} that for any value of $\alpha$, there exists one-cluster coherent state in each layer obeying $|\theta_2^{[i]}-\theta_1^{[i]}|\simeq\pi$, hence $D^{[ii]}\simeq-\alpha$ from Eq.\ref{eqn4}. Also, there exists two populations of inter-layer anti-weights $D^{[ii]}\simeq\alpha$ and $D^{[ii]}\simeq-\alpha$ in the incoherent state, hence Eq.\ref{eqn4} yields $|\theta_2^{[i]}-\theta_1^{[i]}|{=}\arccos(\pm1)=0,\pi$. Hence, the existence of the inhibitory $D^{[ii]}\simeq-\alpha$ gives rise to anti-phase mirror-populations in the two layers in both incoherent and coherent states.

{\em Phase Diagrams:} The data shown in Fig.\ref{figure2} highlight fact that any $\varepsilon\neq0$ is capable of inducing a first-order transition in the system and there exists a marginal difference in the critical coupling strength $\lambda_c^f$ for different values of $\varepsilon$. Fig.\ref{figure5} reports on how the hysteresis width $|r_l^f-r_l^b|$ varies in the $\alpha-\lambda$ space for different values of $\varepsilon$ and for multiplexes composed of two homogeneous (ER-ER) and by one homogeneous (ER) and one heterogeneous (Barab\'asi-Albert (BA)~\cite{Barabasi1999}) networks.
 The ER-ER configuration exhibits a first-order transition for  intermediate and higher values of $\alpha$. Also, the associated hysteresis width for both the layers increases with an increase in $\alpha$. Besides, a slower rate $\varepsilon$ yields a wider hysteresis width for a given value of $\alpha$ as compared to the faster rate $\varepsilon$. For the ER-BA configuration, a rather weak hysteresis is observed for the ER layer in the range $\alpha=0.5-1$, while the BA layer does not feature a first-order transition route to synchronization.
\begin{figure}[t!]
	\centering
		\includegraphics[height=5cm,width=12cm]{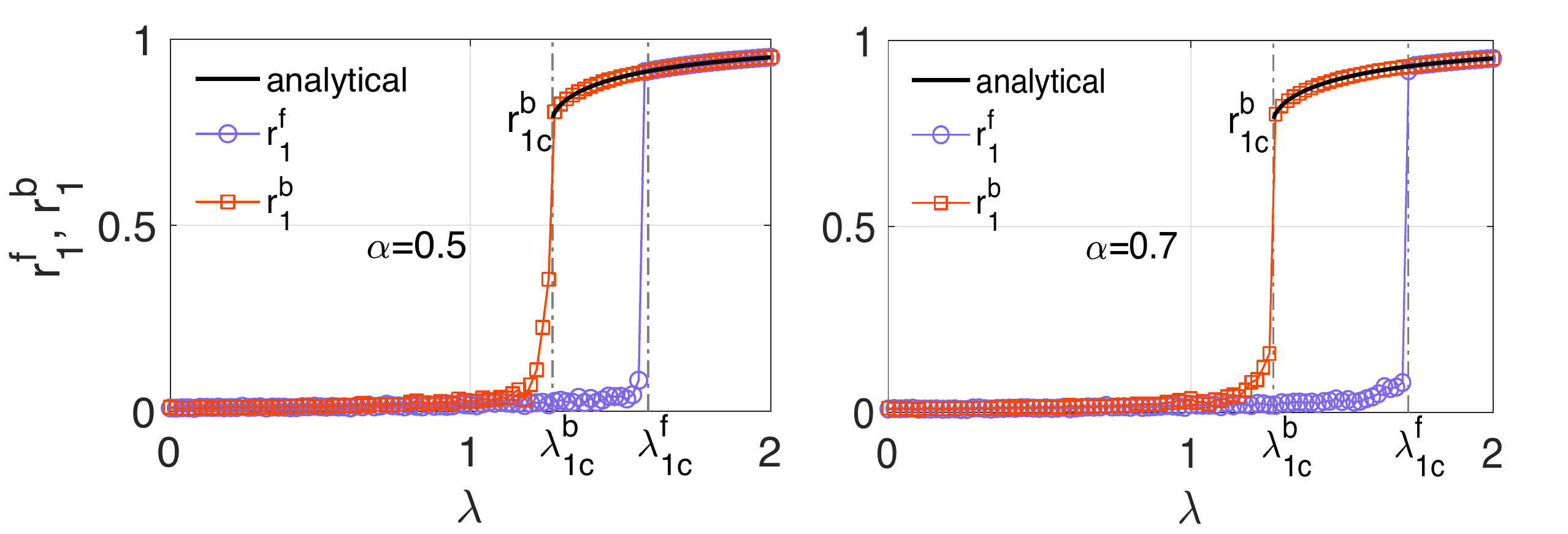}\\
	\vspace{-0.4cm}
	\caption{{\bf Matching with analytical predictions:} Order parameter for a multiplex network comprising two GC layers ($N{=}10,000$ nodes) for $\varepsilon{=}0.1$ and a uniform distribution $g(\omega_l)$ with $\gamma{=}1$. Numerical estimations for backward transition for different values of $\alpha$, and their analytical predictions given by Eqs.\ref{eqn12}, \ref{eqn14} and \ref{eqn17}.}
	\label{figure6}
\end{figure}
\begin{figure}[t!]
	\centering
	\begin{tabular}{cc}
		\includegraphics[height=5cm,width=6cm]{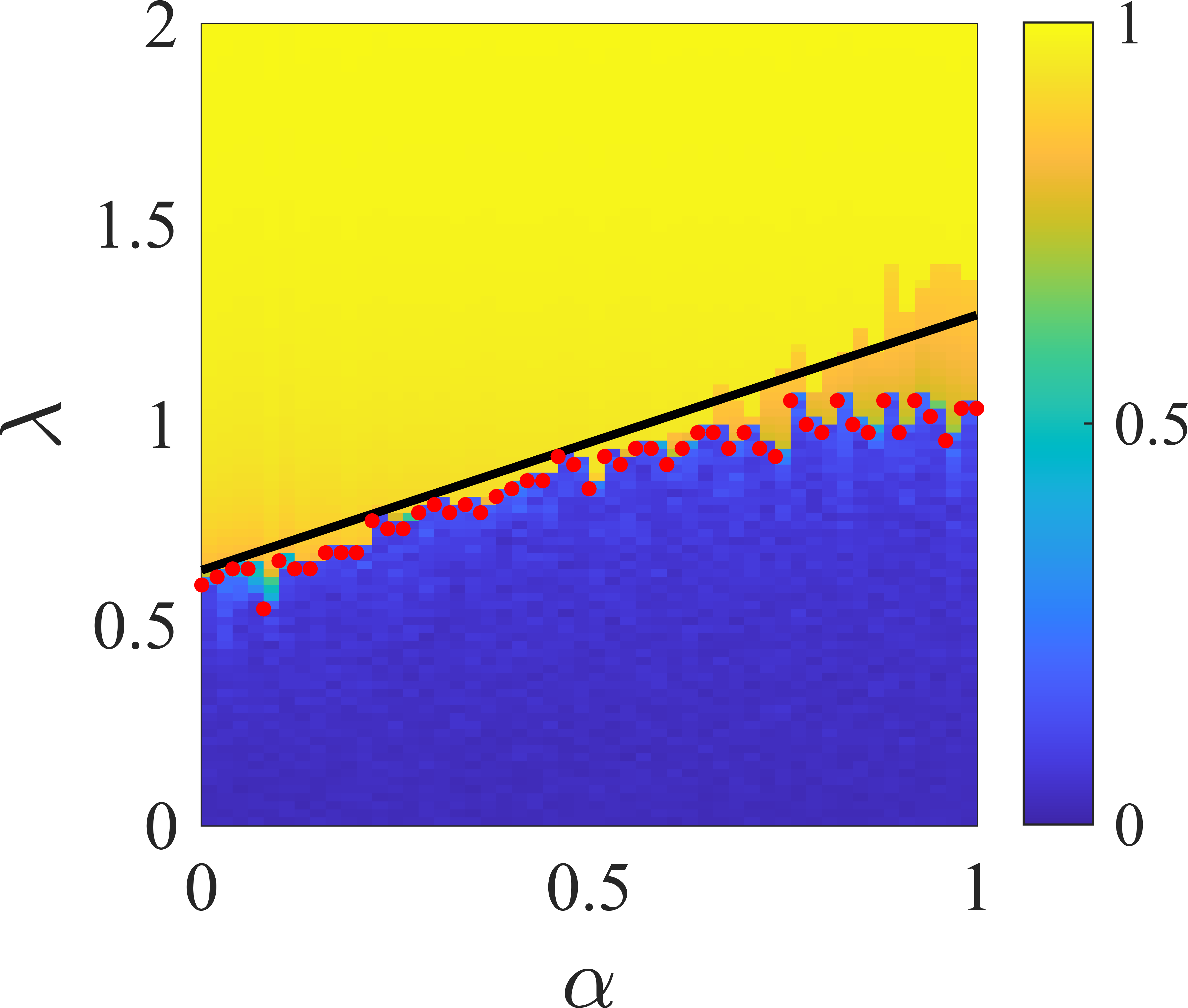}&
		\includegraphics[height=5cm,width=6cm]{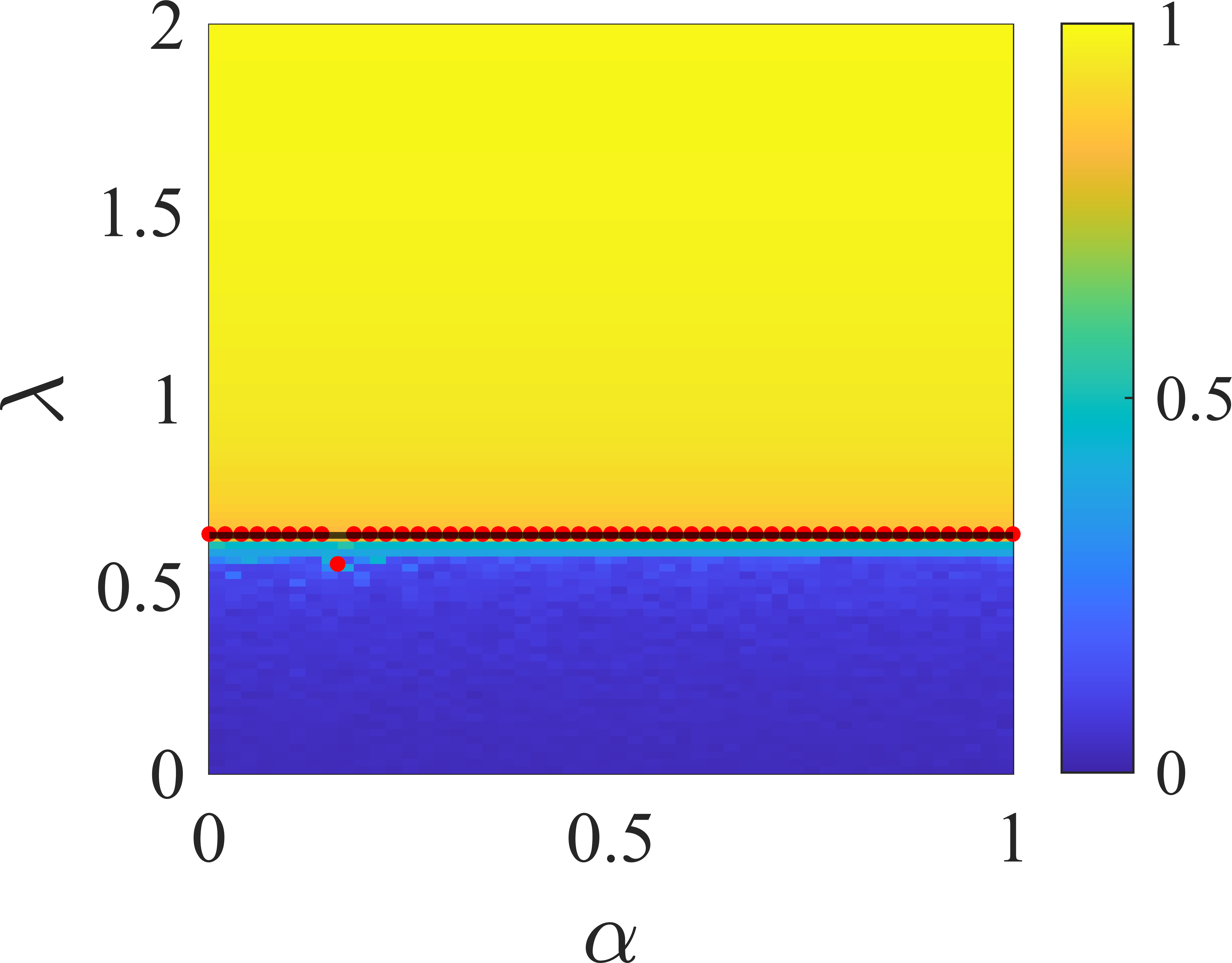}\\
	\end{tabular}{}
	\vspace{-0.4cm}
	\caption{{\bf $\alpha$ dependence of $\lambda_c^f$}: $\lambda_c^f$ and $\lambda_c^b$ (for either layer) as a function of factor $\alpha$ for two GC layers with $\gamma=0.5$ and $\varepsilon=0.1$. Red circle indicates numerical assessment for $\lambda_c^f$ (left panel) and $\lambda_c^b$ (right panel) while black solid line denotes their analytical prediction from Eq.~\ref{eqn14} and Eq.~\ref{eqn17}.}
	\label{figure7}
\end{figure}

\section{Analytical treatment}
To obtain an analytical expression for the order parameter, we take into account, for simplicity, a multiplex network consisting of two globally-connected (GC) layers $l\in[1,2]$ so that $A_l^{ij}=1/N$ in model Eq.\ref{eqn1}. The intrinsic frequencies of the nodes in both GC layers are selected from a uniform distribution $\omega_i=-\gamma+\frac{\gamma}{N}(2i-1)$ in the interval $[-\gamma, \gamma]$.
The order parameter for each layer $l\in[1,2]$ is then defined as
\begin{eqnarray}\label{eqn5}
r_le^{i\psi_l} = \frac{1}{N}\sum_{j=1}^N e^{\imath\theta^{[j]}_l}
\end{eqnarray}
Now Eq.(\ref{eqn1}) can be rewritten in the mean-field form using Eq.(\ref{eqn5}) as
\begin{eqnarray} \label{eqn6}
	\dot{\theta}_{1}^{[i]} = \omega^{[i]}_{1} +  {\lambda}r_1 \sin(\psi_{1}- \theta^{[i]}_{1})
			+ D^{[ii]} \sin(\theta^{[i]}_{2}-\theta^{[i]}_{1}),\nonumber \\
	\dot{\theta}^{[i]}_{2} = \omega^{[i]}_{2} + {\lambda}r_2 \sin(\psi_{2}-\theta^{[i]}_{2}) - D^{[ii]} \sin(\theta^{[i]}_{2}-\theta^{[i]}_{1}).
\end{eqnarray}
In the stationary state $\dot D^{[ii]}=0$, for $\varepsilon\neq0$, thereby Eq.\ref{eqn2} leads to
\begin{eqnarray}
D^{[ii]}=\alpha\cos(\theta^{[i]}_2-\theta^{[i]}_1).
\end{eqnarray}
Hence, evolution of the nodes in the stationary state is ruled by the following self-consistent equations
\begin{eqnarray} \label{eqn7}
	\dot{\theta}_{1}^{[i]} = \omega^{[i]}_{1} +  {\lambda}r_1 \sin(\psi_{1}- \theta^{[i]}_{1})
			+ \frac{\alpha}{2}\sin[2(\theta^{[i]}_{2}-\theta^{[i]}_{1})],\nonumber \\
	\dot{\theta}^{[i]}_{2} = \omega^{[i]}_{2} + {\lambda}r_2 \sin(\psi_{2}-\theta^{[i]}_{2}) - \frac{\alpha}{2}\sin[2(\theta^{[i]}_{2}-\theta^{[i]}_{1})].
\end{eqnarray}

Next, we gather from numerical simulations (see Fig.\ref{figure5}) that the distribution $P(\Delta\theta^{[i]})$ of $\Delta\theta^{[i]}=|\theta_2^{[i]}-\theta_1^{[i]}|$ for the two GC layers in the coherent state ($\lambda>\lambda_c^b$) follows a peaked (unimodal) distribution with its mean at $0$ and standard deviation $\Delta\theta(\lambda)$. Therefore, the inter-layer term in the stationary state can be expressed as $\sigma=\frac{\alpha}{2}\sin(2\Delta\theta)$, and the model Eqs.\ref{eqn6} can be rewritten as
\begin{eqnarray} \label{eqn8}
	\dot{\theta}_{1}^{[i]} = \omega^{[i]}_{1} + \sigma - {\lambda}r_1 \sin(\theta^{[i]}_{1} - \psi_1),\nonumber \\
	\dot{\theta}^{[i]}_{2} = \omega^{[i]}_{2} - \sigma - {\lambda}r_2 \sin(\theta^{[i]}_{2} - \psi_2).
\end{eqnarray}
In this way, stationary $\sigma=\sigma(\alpha,\lambda)$ accounts for the maximum possible inter-layer contribution in the evolution of phases in either layers. When the phases in each layer are locked to their respective mean-fields $\dot\theta^{[i]}_1 = \Omega_1$ and $\dot\theta^{[i]}_2 = \Omega_2$, i.e., $|\omega^{[i]}_1| - \Omega_1+\sigma\leq \lambda r_1$ and $|\omega^{[i]}_2|- \Omega_2 - \sigma\leq \lambda r_2$. Hence, one obtains $\sin(\theta^{[i]}_l - \psi_l) = \frac{\omega^{[i]}_l - \Omega_l \pm \sigma}{\lambda r_l}$,
which leads to the following set of the conditions (referred as CondI) to be satisfied simultaneously by the locked oscillators contributing to $r_l$;\\
CondI:
\begin{eqnarray}
|\omega^i_1| - \Omega_1 + \sigma\leq\lambda r_1, \nonumber \\
|\omega^i_2| - \Omega_2 - \sigma\leq\lambda r_2. \nonumber
\end{eqnarray}
The order parameter in Eq.\ref{eqn5} for the phase-locked oscillators can be expressed as
\begin{equation}
r_l = \frac{1}{N}\sum_{CondI}\cos(\theta^{[j]}_l - \psi_l),
\end{equation}
which can be further simplified as
\begin{eqnarray}\label{eqn10}
 r_l = \frac{1}{N}\sum_{CondI} \sqrt{1-\left(\frac{\omega^{[i]}_l - \Omega_l \pm \sigma}{\lambda r_l}\right)^2},\\
    \Omega_l = \frac{\sum_{CondI} \omega^{[i]}_l}{\sum_{CondI} 1}. \nonumber
\end{eqnarray}
Now, $\Delta\theta\rightarrow 0$ $(\Delta\theta\simeq0.002-0.003)$ for any $\lambda>\lambda_{lc}^b$ corresponding to locked state, hence $\sigma\rightarrow 0$ (see Fig.\ref{figure6}).

{\em Backward Critical Coupling:}
In the continuum limit $N\rightarrow\infty$, the order parameter can be rewritten in its integral form
\begin{equation}\label{eqn11}
 r_l = \int_{CondI} \mathrm{d}\omega_l\ g(\omega_l) \sqrt{1-\left(\frac{\omega_l - \Omega_l \pm\sigma}{\lambda r_l}\right)^2}.
\end{equation}
For a uniform distribution $g(\omega_l)=\frac{1}{2\gamma}$ for $|\omega_l|<\gamma$ and since $\Omega_l \to 0$, the order parameter in Eq.\ref{eqn11} reduces to
\begin{eqnarray}\label{eqn12}
 r_l = \frac{1}{\gamma} \int^{\gamma}_0 \mathrm{d}\omega_l \sqrt{1-\left(\frac{\omega_l \pm\sigma}{\lambda r_l}\right)^2}.
\end{eqnarray}
If $\Delta\theta^{[i]}\sim\pi$ for the pairs of oscillators locked in their respective layers, then $\sigma\to 0$. In such a scenario, we in principal reach to the single layer case for the oscillators locked in their respective layers, thus the order parameter turns into the following:
\begin{equation}\label{eqn13}
r_l = \frac{1}{2} \sqrt{1-\left(\frac{\gamma}{\lambda r_l}\right)^2} + \frac{\lambda r_l}{2 \gamma} \arcsin\left(\frac{\gamma}{\lambda r_l}\right)
\end{equation}
For $\lambda r_l=\gamma$, Eq.~\ref{eqn13} yields $r_{lc}^b{=}\frac{\pi}{4}$. Hence, the backward critical coupling strength is given by
\begin{equation}\label{eqn14}
\lambda_{lc}^b=\frac{\gamma}{r_{lc}^b}=\frac{4\gamma}{\pi}.
\end{equation}

{\em Forward critical coupling:}
From Eq.~\ref{eqn7}, the contribution from inter-layer coupling yields bounds (maximal or minimal) of $\pm\alpha/2$. Moreover, in the incoherent state at $\lambda\rightarrow\lambda_{lc}^f$ ($r_l=0$), the contribution from intra-layer mean-field is negligible. Hence, the evolution of the nodes at $\lambda_{lc}^f$ is driven entirely by the effective critical frequency
\begin{equation}\label{eqn15}
\omega_{lc}^{[i]} \simeq \omega_l^{[i]} \pm \alpha/2.
\end{equation}
So, the critical frequencies at $\lambda_{lc}^f$ are bounded in the effective interval $\omega_{lc}^{[i]}\in(-\gamma-\alpha/2, \gamma+ \alpha/2)$. Hence, the dynamics of the nodes at  $\lambda\rightarrow\lambda_{lc}^f$ can be approximated as
\begin{eqnarray}\label{eqn16}
	\dot{\theta}_{l}^{[i]} = \omega^{[i]}_{lc} +  {\lambda}r_l \sin(\psi_{l}- \theta^{[i]}_{l}).
\end{eqnarray}
Next, following the same methodology adopted for the backward transition, the forward critical threshold is given by
\begin{equation}\label{eqn17}
\lambda_{lc}^f\simeq\frac{4}{\pi}(\gamma+\alpha/2).
\end{equation}

In Fig.~\ref{figure6}, numerical results for the order parameter and the forward and backward thresholds for either GC layers (since both the GC layers synchronizes simultaneously) and their analytical predictions given by Eqs.~\ref{eqn12}, \ref{eqn14} and \ref{eqn17}, are shown for different values of $\alpha$ when $\epsilon=0.1$. The analytical predictions  match fairly well with their numerical estimations. From theoretical and numerical outcomes we deduce that the threshold for first-order de-synchronization does not depend on either $\alpha$ or $\varepsilon$, however the threshold for the onset of first-order transition to synchronization does depend on $\alpha$.
Further, we show the dependence of $\lambda^f_{lc}$ and $\lambda^b_{lc}$ on factor $\alpha$ both numerically as well as analytically as shown in Fig.~\ref{figure7}. The $\alpha-\lambda$ phase plot unveils that $\lambda_{lc}^b$ remains independent of $\alpha$ and fixed to $\frac{4\gamma}{\pi}$. However, $\lambda^f_{lc}$ does show dependence on $\alpha$ closely following Eq.~\ref{eqn17}.

\section{Robustness against network size $N$ and structure}
\begin{figure}[t!]
	\centering
	\includegraphics[height=8cm,width=12cm]{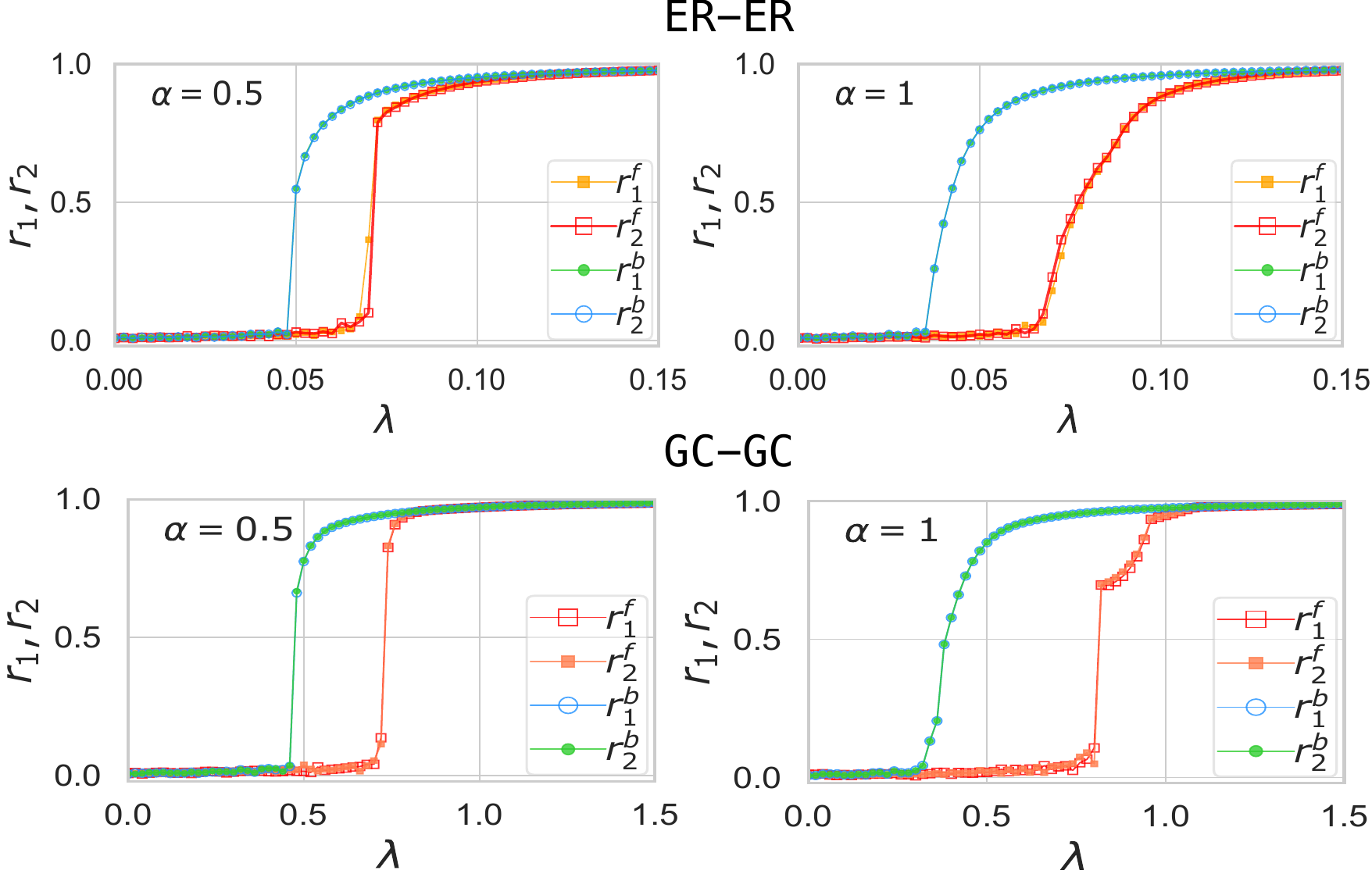}\\
	\vspace{-0.5cm}
	\caption{{\bf For larger size networks:} Synchronization profile ($r_l-\lambda$) for larger network size ($N=10000$) of multiplexed ER-ER layers and GC-GC layers. Other parameters are $\langle k_1\rangle=\langle k_2\rangle=10$, $\epsilon=0.1$ and $\gamma=0.5$.}
	\label{figure8}
\end{figure}
We explored the phenomena of ES as a consequence of adaptive multiplexing for larger size as well as for different network architecture. Numerical results for larger sizes $N=10,000$ of multiplex networks composed of two different pairs of homogeneous topology, namely ER-ER and GC-GC network are shown in Fig.~\ref{figure8}. Numerical results for the order parameter corresponding to different values of $\alpha$ for large size $N$ are consistent with those obtained for rather small network size $N=1,000$. Hence, the occurrence of ES as an outcome of inter-layer adaptation is robust against networks size $N$ and homogeneous network topology of the multiplex network. Nevertheless, the employed inter-layer Hebbian adaptive mechanism is incapable in triggering ES transition in heterogeneous topology for a multiplexed layer as can be observed for BA layer in the phase plots for ER-BA multiplex configuration in Fig.~\ref{figure5}.

\section{Conclusion}
We studied the adaptive evolution of connection weights of inter-layer links in multiplex networks. The connection weight between a pair of interconnected nodes is strengthened if they are in phase and weakened if they are out of phase. Such an Hebbian learning rule plays an important role in shaping the dynamics of the phases, as well as inter-layer links control, in turn, intra-layer synchronization. In the asynchronous state of each layer, the existing Hebbian learning adaptation divides the inter-layer weight population into two almost equal groups corresponding to steady states $\alpha$ and $-\alpha$, yielding a phase-difference of $0$ and $\pi$, respectively, between the interconnected nodes. It is the almost half anti-weight population $-\alpha$ the one which generates frustration at its end interconnected nodes and induces a first-order like transition. In the synchronous state of each layer, the Hebbian learning adaptation pulls the entire inter-layer link population into a single group corresponding to a $-\alpha$ steady state with the interconnected nodes maintaining a phase-difference of $\pi$. It is also unveiled that Hebbian learning rule provides a great amount of control over the abruptness (of the first-order transition) and the size of the associated hysteresis by means of the amplification factor $\alpha$, however the effect of the learning rate $\varepsilon$ is weak. The backward critical $\lambda_c^b$ is independent on both $\alpha$ and $\varepsilon$, while $\lambda_c^f$ is shown to have explicit dependence on $\alpha$. The numerical estimation of both $\lambda_c^f$ and $\lambda_c^b$ have shown to fall in good agreement with their respective analytical predictions. The proposed scheme of inter-layer Hebbian adaptation is capable to bringing about first-order transition in homogeneous multiplexed layers. Phase diagrams for hysteresis width $|r_l^f-r_l^b|$ in $\alpha-\lambda$ space for different rates $\varepsilon$ are provided for multiplex networks with different combinations of topology.

\appendix
\section{$\omega_1^{[i]}=\omega_2^{[i]}$ case}
\begin{figure}[t!]
	\centering
	\begin{tabular}{cccc}
	\includegraphics[height=12cm,width=16cm]{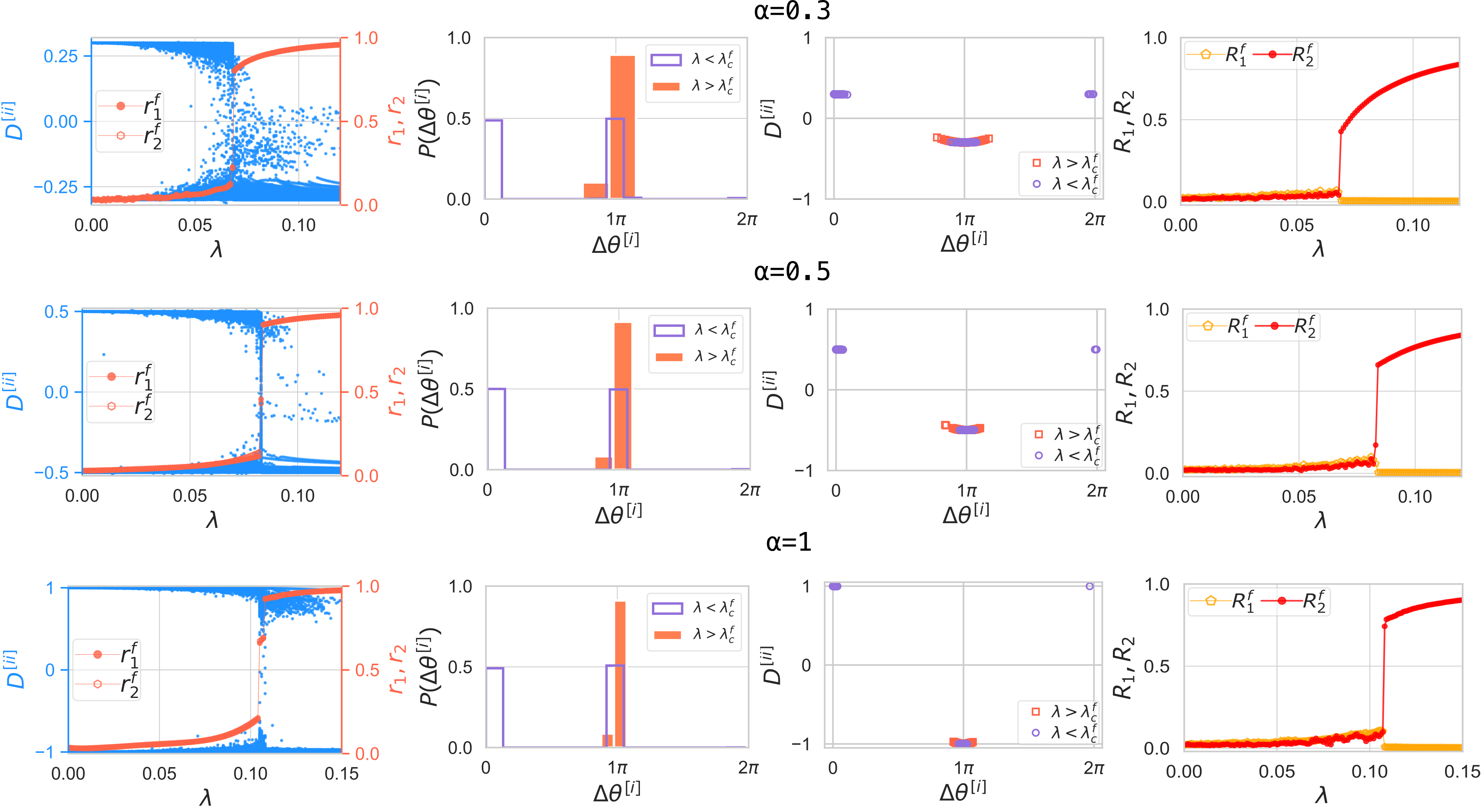}\\
	\end{tabular}{}
	\vspace{-0.5cm}
	\caption{{\bf Mechanism behind ES for $\omega_1^{[i]}=\omega_2^{[i]}$:} $D^{[ii]}$ as a function of $\lambda$, the distribution $P(\Delta\theta^{[i]})$ of $\Delta\theta^{[i]}=|\theta_2^{[i]}-\theta_1^{[i]}|$ belonging to the incoherent and coherent state, $D^{[ii]}$ against $\Delta\theta^{[i]}$, and $R_1$ and $R_2$ order parameter as a function of $\lambda$ for the multiplex network comprising two ER layers. Results are produced for $N{=}1000$ nodes, $\varepsilon=0.1$ and $\gamma=0.5$ and different values of $\alpha$.}
	\label{figure9}
\end{figure}

The microscopic dynamics behind the origin of ES is simplified further when taking into account $\omega_1^{[i]}=\omega_2^{[i]}$ for multiplex networks made of two ER layers. The first column in Fig.~\ref{figure9} shows the absence of neutral cluster around $D^{[ii]}\rightarrow0$ which is present for the case of $\omega_1^{[i]}\neq\omega_2^{[i]}$, hence the identical frequencies of the interconnected nodes leads to only two pure anti-weight states, namely $D^{ii}\simeq\alpha$ and $D^{ii}\simeq-\alpha$. And for that matter, only two equal sized clusters are obtained following $|\theta_2^{[i]}-\theta_1^{[i]}|=0$ and $\pi$ in the incoherent state (see second column). Anyhow, one single peak is obtained in the coherent state.
In the third column, the behavior of the final states of $D^{[ii]}$ against $\Delta\theta^{[i]}$ affirms the fact that for any $\alpha$ in the coherent state, a single phase-cluster with $|\theta_2^{[i]}-\theta_1^{[i]}|=\pi$ originates from the cloud of $D^{[ii]}\rightarrow-\alpha$. In the incoherent state, one-half population of the inter-layer link experiencing $D^{[ii]}\rightarrow\alpha$ gives birth to a phase-cluster with $\theta_2^{[i]}=\theta_1^{[i]}$ in each layer while the other-half population experiencing $-\alpha$ leads to phase-cluster with $|\theta_2^{[i]}-\theta_1^{[i]}|=\pi$ in each layer. The fourth column infers that the abrupt jump in $R_2$ at the onset of synchronization shows the existence of anti-phase layers. Also, the special case of $\omega_1^{[i]}=\omega_2^{[i]}$ shows a handsome jump size in $R_2$ for any value of $\alpha$ as that for the case of $\omega_1^{[i]}\neq\omega_2^{[i]}$.

\ack
SJ thanks CSIR grant 25(0293)/18/EMR-II for financial support. ADK acknowledges Govt of India CSIR grant 25(0293)/18/EMR-II for RA fellowship.

\section*{References}
\providecommand{\newblock}{}

\end{document}